\documentclass[12pt]{article}
\usepackage{graphicx}

\usepackage{times}

\newcommand{\aap}{{Astron. Astrophys.}}
\newcommand{\apj}{{Astrophys. J.}}
\newcommand{\apjl}{{Astrophys. J. Lett.}}

\newcommand{\grl}{{Geophys. Res. Lett.}}

\newcommand{\ssr}{{Space Sci. Rev.}}

\newenvironment{sciabstract}{%
\begin{quote} \bf}
{\end{quote}}

\newcounter{lastnote}

\title{The crucial role of surface magnetic fields for the solar dynamo}

\author
{Robert Cameron$^{1\ast}$ and Manfred Sch{\"u}ssler$^{1}$\\
\\
\normalsize{$^{1}$Max-Planck-Institut f\"ur Sonnensystemforschung,
   G{\"o}ttingen 37077, Germany}
\\
\normalsize{$^\ast$To whom correspondence should be addressed; 
            E-mail:  cameron@mps.mpg.de}
}

\date{}

\begin{document}

\baselineskip24pt

\maketitle

\begin{sciabstract}

{Sunspots and the plethora of other phenomena occuring in the course of
the 11-year cycle of solar activity are a consequence of the emergence
of magnetic flux at the solar surface. The observed orientations of
bipolar sunspot groups imply that they originate from toroidal
(azimuthally orientated) magnetic flux in the convective envelope of the
Sun.  We show that the net toroidal magnetic flux generated by
differential rotation within a hemisphere of the convection zone is
determined by the emerged magnetic flux at the solar surface and thus
can be calculated from the observed magnetic field distribution. The
main source of the toroidal flux is the roughly dipolar surface magnetic
field at the polar caps, which peaks around the minima of the activity
cycle.}

\end{sciabstract}

The basic concept for the large-scale solar dynamo involves a cycle
during which the poloidal field and the toroidal field are mutually
generated by one another ({\it 1, 2}). The winding
of the poloidal field by differential rotation creates a toroidal field.
A reversed poloidal field results from the formation of magnetic loops
in the toroidal field, which become twisted by the Coriolis force due to
solar rotation.  In turn, the reversed poloidal field then becomes the
source of a reversed toroidal field. In this way, the 11-year cycle of
solar activity is connected to a 22-yr cycle of magnetic polarity.

Hale et al. ({\it 3}) discovered that the magnetic
orientations of the eastward and westward parts of bipolar sunspot
groups in one solar hemisphere are the same during an 11-year cycle and
opposite in the other hemisphere. This implies that the sunspot groups
originate from a toroidal field of fixed orientation during a cycle.
Toroidal flux of the opposite polarity would lead to sunspot groups
violating Hale's law. Since only a small minority of the
sunspot groups are actually observed to violate this
rule ({\it 4}), opposite-polarity toroidal field is
largely irrelevant as a source of sunspot groups.  In other words, it is
the hemispheric net toroidal magnetic flux given by the azimuthal
average of the toroidal field that is relevant for the formation of
sunspot groups.

Here we use a simple method based on Stokes' theorem to show that 
the emerged surface fields determine the net toroidal flux generated
by differential rotation in a solar hemisphere.  The time evolution of
the net toroidal flux in the convection zone can thus be calculated
using only observed quantities (differential rotation and field
distribution at the surface). We compare the resulting net
toroidal flux with the observed large-scale unsigned surface flux and
find that they vary in a similar manner.

We consider spherical polar coordinates, $(r,\theta,\phi)$, and the
azimuthally averaged induction equation of magnetohydrodynamics,
\begin{equation}
\frac{\partial {\bf{B}}}{\partial t}=
     \nabla \times \left({\bf{U}}\times {\bf{B}}
     + \langle {\bf{u}}\times {\bf{b}} \rangle  
     -\eta \nabla \times {\bf{B}} \right),
\label{eq:induction}
\end{equation}
where ${\bf B}(r,\theta)$ and ${\bf U}(r,\theta)$ are the
$\phi$-averaged magnetic field and plasma velocity,
respectively, and $\eta$ is the magnetic diffusivity. Angular brackets
indicate the azimuthal average. The term
$\langle{\bf{u}}\times{\bf{b}}\rangle$ denotes the correlation of the
fluctuating quantities with respect to the azimuthal averages, which
gives rise to the $\alpha$-effect and to enhanced (`turbulent') magnetic
diffusivity ({\it 5}).

\begin{figure}
\begin{center}
\includegraphics[scale=0.3]{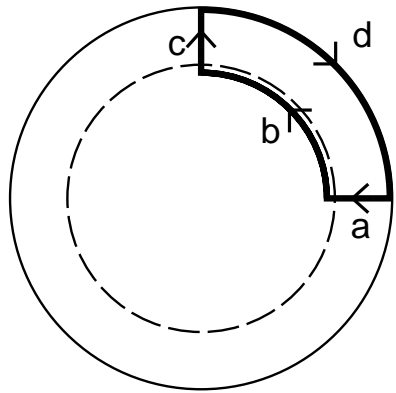}
\caption{Integration contour for the application of Stokes' theorem.
The contour (thick solid line) on a meridional plane of the Sun is used
to calculate the net toroidal flux in the northern hemisphere generated
by the action of differential rotation on the poloidal field.  The thin
solid line represents the solar surface, the dashed line the bottom of
the convection zone. The rotation poles are at the top and bottom.  The
contour consists of a radial segment in the equatorial plane (a), a
circular arc slightly below the bottom of the convection zone (b), a
part along the axis of rotation (c), and the solar surface (d).}
\label{fig:ill}
\end{center}
\end{figure}

We define the contour $\delta \Sigma$ enclosing the area $\Sigma$ in a
meridional plane of the Sun as shown in Fig.~\ref{fig:ill}. The
direction of the contour is chosen such that the vectorial surface
element of $\Sigma$ points into the direction of positive azimuthal
field, $B_\phi$.  Applying Stokes' theorem to the integral of the
induction equation over $\Sigma$ yields the time derivative of the net
toroidal flux, $\Phi^{\rm N}_{\rm tor}$, in the northern hemisphere of
the convection zone,
\begin{equation}
          \frac{{\rm d} \Phi^{\rm N}_{\rm tor}}{{\rm d} t} = 
\frac{{\rm d}}{{\rm d} t} \left(\int_{\Sigma} B_\phi\mathrm{d}{S}\right)
         = \int_{\delta \Sigma} {\left({\bf{U}}\times {\bf{B}}
         + \langle{\bf{u}}\times{\bf{b}}\rangle  
         -\eta \nabla \times {\bf{B}} \right)} \cdot \mathrm{d}{\bf l}\,,
\label{eq:btor_gen}
\end{equation}
where $\mathrm{d}{S}$ is the surface element of $\Sigma$ and
$\mathrm{d}{\bf l}$ is the line element along ${\delta \Sigma}$. An
analoguous procedure provides the net toroidal flux in the southern
hemisphere, $\Phi^{\rm S}_{\rm tor}$.

Rotation is by far the dominant component of the azimuthally averaged
velocity, so that we write ${\bf U}=U_\phi\,\mbox{\boldmath$\hat{\phi}$}
= (\Omega \, r \sin\theta)\mbox{\boldmath$\hat{\phi}$}$, where
$\Omega(r,\theta)$ is the angular velocity and
\mbox{\boldmath$\hat{\phi}$} the unit vector in the azimuthal direction.
The effect of
$\int_{\delta\Sigma}\langle{\bf{u}}\times{\bf{b}}\rangle\cdot\mathrm{d}{\bf l}$ reduces to
that of the turbulent magnetic diffusivity, $\eta_{\rm t}$, since
the contribution of the $\alpha$-effect to the generation of the
toroidal field can be neglected against that of differential
rotation ({\it 1}). With $\eta_{\rm t}\gg\eta$ we
thus obtain
\begin{equation}
          \frac{{\rm d} \Phi^{\rm N}_{\rm tor}}{{\rm d} t} 
          = \int_{\delta \Sigma} {\left({\bf{U}}\times {\bf{B}}
         -\eta_{\rm t}\nabla\times{\bf{B}}\right)} \cdot \mathrm{d}{\bf l}\,.
\label{eq:btor_gen_2}
\end{equation}

Guided by empirical results from helioseismology ({\it 6, 7}),
we take $\Omega$ to be independent of $r$ in the equatorial plane
throughout the convection zone ({\it 8}), i.e., $\Omega(r,\pi/2) =
\Omega_{\rm eq}$. This allows us to work in a reference frame rotating
with angular velocity $\Omega_{\rm eq}$, for which $U_\phi=0$ in the
equatorial plane. We can further assume that the magnetic field does not
penetrate the low-diffusivity radiative zone below the convection
zone. Together with $U_\phi=0$ along the rotational axis, these
assumptions imply that only the surface segment (d) in
Fig.~\ref{fig:ill} contributes to the line integral of ${\bf{U}}
\times{\bf{B}}$ along the contour $\delta \Sigma$. We obtain
\begin{equation}
           \int_{\delta \Sigma} \left({\bf{U}}\times {\bf{B}}\right) 
           \cdot  \mathrm{d}{\bf l}= \int_{0}^{\pi/2} U_{\phi}B_r
           R_{\odot}  \mathrm{d} \theta =
	   \int_{0}^{1} (\Omega-\Omega_{\rm eq}) B_r 
           R_{\odot}^2  \,\mathrm{d}(\cos\theta) \,,
\label{eq:btor_induction}
\end{equation}
where $U_\phi$, $\Omega$, and $B_r$ are to be taken at the solar
surface, $r=R_\odot$.  This shows that the net toroidal flux
generated in the convection zone by the action of differential rotation
is determined by the poloidal field threading the solar surface.
Any additional poloidal flux that is fully contained within
the convection zone would lead to equal amounts of East-West and
West-East orientated toroidal flux, which do not contribute to the net
toroidal flux required by Hale's law.

The diffusion term in Eq.~(\ref{eq:btor_gen_2}) is most relevant along
the rotational axis, where toroidal flux can be destroyed, and also in
the equatorial segment of $\delta\Sigma$, where flux can be transported
between the hemispheres.  These processes, which decrease the net
magnetic flux, are difficult to quantify. In order to estimate their
effect, we approximate them with an exponential decay term with
$e$-folding time $\tau$.  Altogether we then obtain
\begin{equation}
   \frac{{\rm d} \Phi^{\rm N}_{\rm tor}}{{\rm d} t}= 
           \int_{0}^{1} (\Omega-\Omega_{\rm eq}) B_r 
           R_{\odot}^2  \,\mathrm{d}(\cos\theta) 
           -\frac{\Phi^{\rm N}_{\rm tor}}{\tau},
\label{eq:btor_end}
\end{equation}
and an analoguous equation for the net toroidal flux generation in the
southern hemisphere of the convection zone.

\begin{figure}
\begin{center}
\includegraphics[scale=0.6]{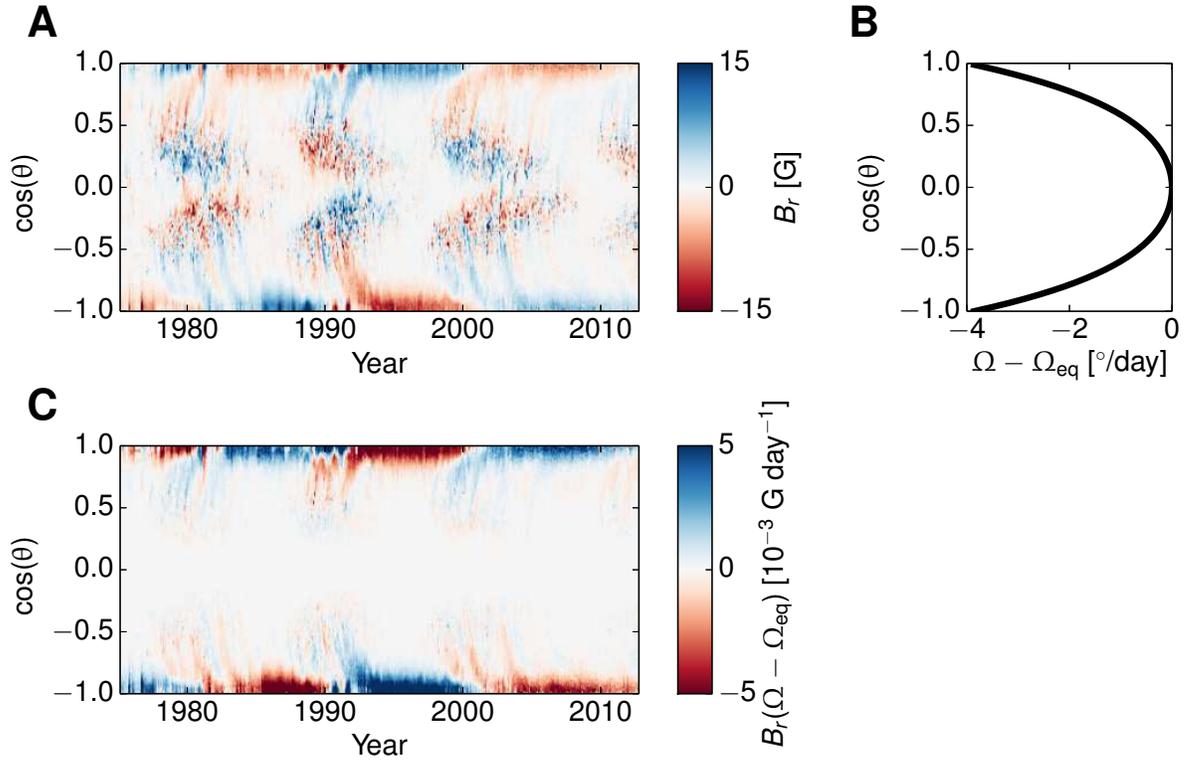}
\caption{A: Azimuthally averaged radial surface field from observed
synoptic magnetograms as a function of $\cos\theta$ and time. B: Solar
surface differential rotation relative to the equator as a function of
$\cos\theta$.  C: Map of the quantity $(\Omega-\Omega_{\rm eq}) B_r$,
representing the source term for the generation of net toroidal flux in
Eq.~(\ref{eq:btor_induction}), as a function of $\cos\theta$ and time.}
\label{fig:mbf}
\end{center}
\end{figure}

To evaluate the inductive part of Eq.~(\ref{eq:btor_end}) we use the
synoptic magnetograms from the Kitt Peak National Observatory from 1975
to the present ({\it 9}).  The azimuthally averaged radial component
of the magnetic field, $B_r$, as a function of time and $\cos\theta$ is
shown in Fig.~\ref{fig:mbf}A, where $\cos\theta=0$ represents the
equator and $\cos\theta=\pm1$ the poles.  The profile of the surface
differential rotation taken from ({\it 10}),
\begin{equation}
\Omega-\Omega_{\rm eq}=-2.3 \cos^2\theta-1.62 \cos^4\theta 
               \;\; [^{\circ}/{\rm day}]\,,
\label{eqn:DR}
\end{equation}
given in Fig.~\ref{fig:mbf}B is combined with $B_r$ to obtain the
quantitity $(\Omega-\Omega_{\rm eq})B_r$ (Fig.~\ref{fig:mbf}C), which
determines the generation of toroidal flux by differential rotation in
Eq.~(\ref{eq:btor_end}). It is dominated by the contribution from the
polar regions. 

\begin{figure}
\begin{center}
\includegraphics[scale=0.8]{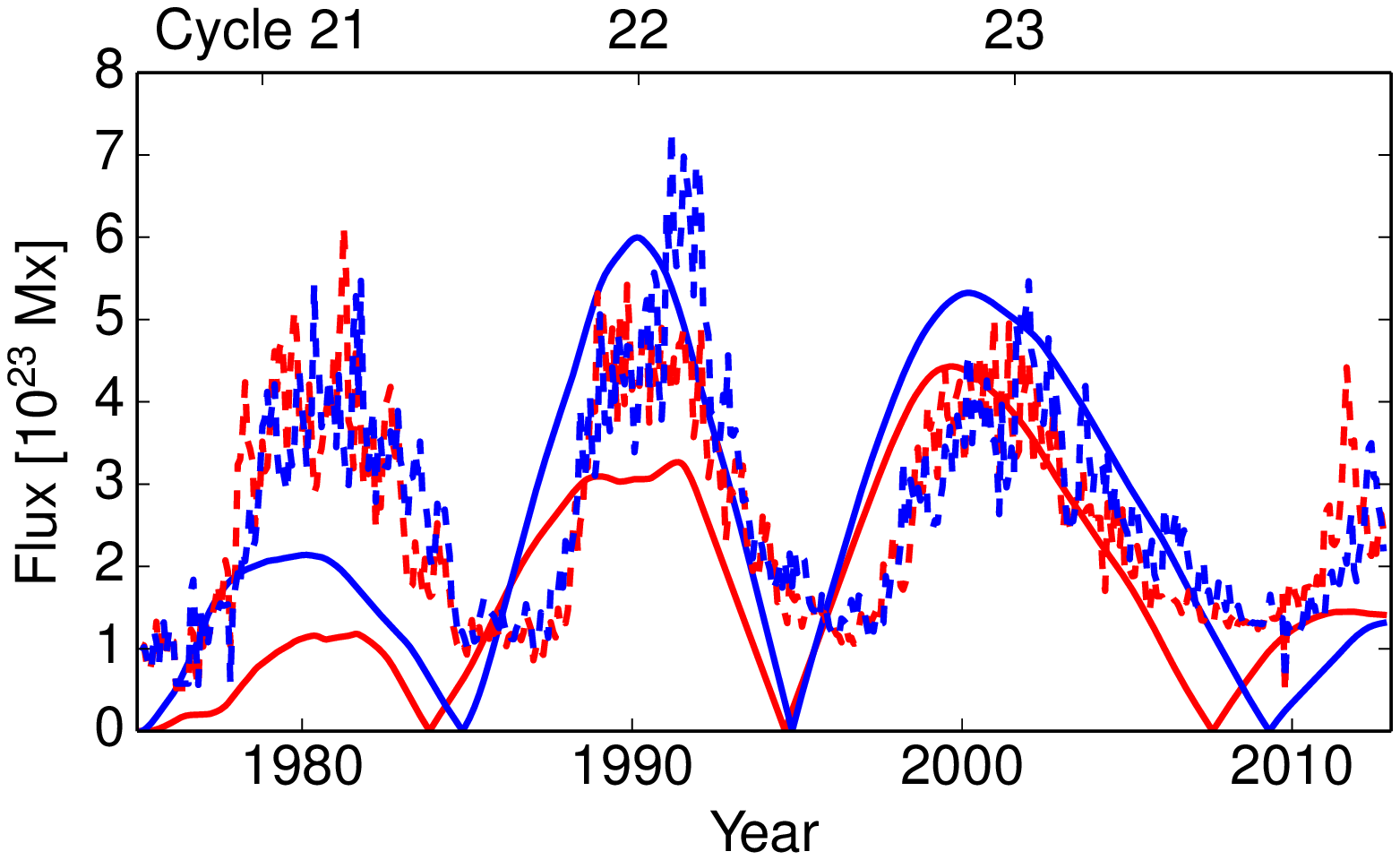}
\caption{Observed magnetic flux at the solar surface and calculated net
  toroidal flux. Hemispheric total unsigned surface flux (dashed
  lines) from synoptic magnetograms  and the modulus of the net toroidal
  flux (solid lines) are given for the last three solar cycles. Red lines
  refer to the northern and blue lines to the southern hemisphere.}
\label{fig:int}
\end{center}
\end{figure}

We calculated the time integral of the inductive part of
Eq.~(\ref{eq:btor_end}) as well as its counterpart for the southern
hemisphere to obtain the net toroidal flux in both hemispheres as a
function of time. The integration begins with zero flux in February
1975, the starting time of the synoptic observations. This is near solar
activity minimum, during which time we expect the toroidal flux to
change sign.  The result given in Fig.~\ref{fig:int} shows that the
modulus of the toroidal flux generated from the polar fields reaches
peak values of the order of $1-6\times10^{23}$~Mx per hemisphere during
recent activity cycles. Note that the net flux generated for the new
cycle first has to cancel the opposite-polarity flux from the old cycle,
so that it reaches its peak value around activity maximum of the new
cycle.

The exponential decay term in Eq.~(\ref{eq:btor_end}) mainly leads to a
phase shift of the time evolution: the sign reversals and the maximum
values of the toroidal flux occur earlier since the flux from
the previous cycle is continuously reduced by the decay. At the same
time, the newly generated flux is also subject to the decay, so that the
amplitude of the toroidal flux is only weakly affected: even for the
extreme case of $\tau=4\,$years, the peak values for cycles 22 and 23
are reduced by at most 20--30\%. Plots analoguous to
Figure~\ref{fig:int} for various values of $\tau$ are provided in the
online supplementary material.

Is the inferred amount of net toroidal flux sufficient to provide the
magnetic flux actually emerging in the form of bigger bipolar magnetic
regions and sunspot groups at the solar surface? In order to estimate
their combined fluxes, which provide a rough measure of the subsurface
toroidal flux, we calculated the integrated unsigned radial surface
flux, $\Phi_{\rm U}=\int|B_r|{\rm d}A$ (where ${\rm d}A$ represents the
solar surface element) for each hemisphere on the basis of the Kitt Peak
synoptic magnetograms. Because the spatial resolution element of the
synoptic magnetograms is on average about $12\times 12$ Mm$^{2}$ on the
solar surface, $\Phi_{\rm U}$ represents the large-scale flux
originating from the emergence of bigger bipolar magnetic
regions. Fig.~\ref{fig:int} shows that the time evolution of $\Phi_{\rm
U}$ for both hemispheres is very similar to that of the inferred
toroidal flux, both with respect to phase as well as to amplitude
({\it 11}).  A short decay time, $\tau$, would lead to a significant
phase shift, suggesting that the decay term in Eq.~(\ref{eq:btor_end})
has no strong effect on the net flux.  Fig.~\ref{fig:int} indicates that
the toroidal flux generated from the polar fields is sufficient to
explain the observed time profile and amount of flux emerging in the
resolved bipolar regions and sunspot groups in the course of the solar
cycle. Since there is probably considerable randomness in the flux
emergence process, leading to elements of toroidal flux emerging more
than once at various longitudes or not emerging at all, we cannot expect
more than an order-of-magnitude agreement.

Our results demonstrate that the emerged magnetic flux and particularly
the polar fields are by far the dominating source of the net toroidal
flux in the convection zone, from which the sunspot groups of the
subsequent cycle originate. The solenoidality of the magnetic field
means that the flux associated with the polar fields threads through the
Sun's convection zone, where it is wound up by differential rotation to
generate toroidal magnetic flux. We thus confirm the conjecture of
H.W. Babcock ({\it 12}) that the observed polar fields represent
the poloidal field source for the subsurface toroidal field.

As also first suggested by Babcock, the net axial dipole moment
represented by the polar fields results from the North-South dipole
moments contributed by the individual sunspot groups and bipolar
magnetic regions at the surface as a result of their systematic tilt
with respect to the East-West direction ({\it 3}). The tilt probably
originates in one way or another from rotation via the Coriolis force:
either by providing helicity to convective flows bringing magnetic flux
to the surface or by twisting buoyantly rising flux loops. The concept
of Babcock was further developed by Leighton ({\it 13}), who
introduced the notion of surface flux transport for the buildup of the
polar fields in connection with the dynamo process.  More recently,
surface flux transport models successfully reproduced the observed
evolution of the surface fields and, in particular, the polar fields on
the basis of the observed records of sunspot groups as flux
input ({\it 14--18}). This implies that the tilt of the larger bipolar magnetic
regions determines the polar fields. Small bipolar regions and
small-scale correlations are irrelevant in this respect.  Together with
the results shown here, this establishes the key part of the surface
fields in the solar dynamo process and thus corroborate the basic dynamo
concept of Babcock ({\it 12}) and Leighton ({\it 13}).

The key role played by the polar fields for the generation of toroidal
flux explains the strong empirical correlation between the strength of
the polar field ({\it 19}) and the Sun's open
flux ({\it 20}) around activity minimum with the number of sunspots
of the subsequent activity cycle, which can be taken as a proxy for
the underlying toroidal flux. Although the correlation is not perfect,
which can be ascribed to randomness associated with the flux
emergence process, it provides the best available method to predict
the strength of the next cycle ({\it 21, 22}).
Our results put this method on a firm physical basis.

\noindent{\bf Acknowledgments}: The authors are grateful to Matthias
Rempel for enlightening discussions that led to significant improvement
of the manuscript. This work was
carried out in the context of Deutsche Forschungsgemeinschaft
SFB 963 “Astrophysical Flow Instabilities and Turbulence”
(Project A16).

\newpage

\begin{thebibliography}{10}
\bibitem{note3}{The low values of the toroidal flux for cycle 21 are
  presumably due to the poorer quality of the earlier synoptic
  magnetograms before significant improvements of the instrumentation
  were implemented.}
\end{thebibliography}
\end{document}